\documentclass{eas}
\usepackage{graphicx}
\newcommand{\kms}{kms$^{-1}$} 
\newcommand{\vsini}{$v\sin i$} 
\newcommand{\msunyr}{M$_\odot$yr$^{-1}$}

\newcommand{\aap}{A\&A}
\newcommand{\apj}{ApJ}
\newcommand{\apjs}{ApJS}
\newcommand{\apjl}{ApJL}
\newcommand{\pasp}{PASP}
\newcommand{\mnras}{MNRAS}
\newcommand{\aj}{AJ}

\newcommand{\aapr}{A\&AR}
\newcommand{\solphys}{Sol. Phys.}

\begin{document}

\title{Observational studies of stellar rotation} 
\author{J. Bouvier}\address{UJF-Grenoble 1 / CNRS-INSU, Institut de
  Plan\'etologie et d'Astrophysique de Grenoble (IPAG) UMR 5274,
  Grenoble, F-38041, France }
%
%
\begin{abstract}
  This course reviews the rotational properties of non-degenerate
  stars as observed from the protostellar stage to the end of the main
  sequence. It includes an introduction to the various observational
  techniques used to measure stellar rotation. Angular momentum
  evolution models developed over the mass range from the substellar
  domain to high-mass stars are briefly discussed.  \end{abstract}
\maketitle
\section{Introduction}

The angular momentum content of a star at birth impacts on most of its
subsequent evolution (e.g. Ekstr\"om et al. 2012). The star's
instantaneous spin rate and/or on its rotational history plays a
central role in various processes, such as dynamo-driven magnetic
activity, mass outflows and galactic yields, surface chemical
abundances, internal flows and overall structure, and it may as well
influences the planetary formation and migration processes. It is
therefore of prime importance to understand the origin and evolution
of stellar angular momentum, indeed one of the most challenging issues
of modern stellar physics. Conversely, the evolution of stellar spin
rate is governed by fundamental processes operating in the stellar
interior and at the interface between the star and its immediate
surroundings. The measurement of stellar rotation at various
evolutionary stages and over a wide mass range thus provides a
powerful means to probe these processes.

In this introductory course, an overview of the rotational properties
of stars and of angular momentum evolution models is provided. In
Section~\ref{tech}, various techniques used to measure stellar
rotation are described. In Section~\ref{lowmass}, the rotational
properties of solar-type and low-mass stars are reviewed. Angular
momentum evolution models developed for low-mass stars are discussed in
Section~\ref{models}. Finally, the rotational properties of
intermediate-mass and massive stars are briefly outlined in
Section~\ref{highmass}.

\section{Measurement techniques}
\label{tech}

\begin{figure*}
   \centering
  \includegraphics[angle=0,width=\hsize]{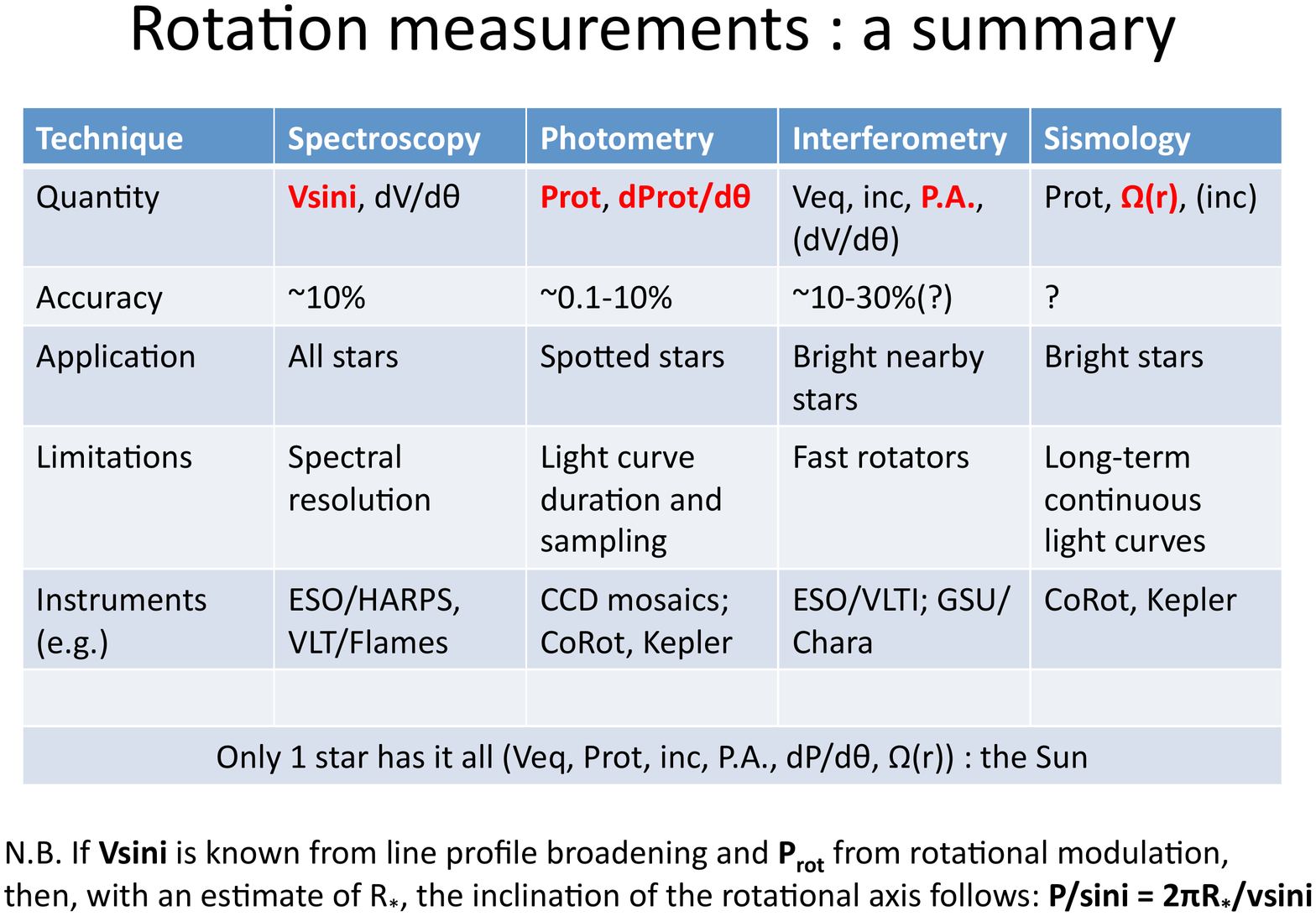}
 \caption{A summary of measurement techniques used to derive
   the rotational properties of non-degenerated stars.}
              \label{meas}%
    \end{figure*}

Stellar rotation can be measured through a variety of techniques. I
illustrate here some of the most commonly applied ones to measure the
rotation rates of non-degenerated objects. The various techniques are
summarized in Figure~\ref{meas}. 

\subsection{Spectroscopy}

Capt. Abney (1877) was apparently the first to consider the effect
rotation would have on a stellar spectrum. He suggested that Doppler
broadening of the photospheric line profiles should occur, as the
light from the rotating surface goes through the entrance slit of the
spectrograph. For a star with a linear equatorial velocity $V_{eq}$,
the spectral broadening of photospheric lines amounts to
$\Delta\lambda_L=(\lambda/c)\cdot V_{eq}\cdot \sin i$, where $i$ is
the inclination angle between the line of sight and the rotation
axis. A star seen pole-on ($i$=0) exhibits no Doppler broadening,
while a direct measurement of $V_{eq}$ is obtained for an equator-on
star ($i$=90$\deg$). The isorotation locus on the stellar disk, i.e.,
points of the stellar surface having the same projected velocity,
follows vertical stripes parallel to the rotational axis, whose
wavelength shift is given by $\Delta\lambda = (\lambda/c)\cdot
V_{eq}\sin i \cdot cos(l) \cdot sin (L)$, where $l$ and $L$ are
respectively the latitude and longitude of a point at the stellar
surface.

The integrated line profile of a rotating star is the sum of the
intrinsic line profiles of all points on the stellar disk affected by
their respective Doppler shifts. To first order, it can be described as
the convolution product of the intrinsic, non-rotating line profile
with a ``broadening'' function given by (cf. Carroll 1933; Gray 1973):

$$G(\lambda) = {2(1-\epsilon)[1-(\Delta\lambda/\Delta\lambda_L)^2]^{1/2} +
{1\over 2} \pi\epsilon )[1-(\Delta\lambda/\Delta\lambda_L)^2]\over\pi
\Delta\lambda_L(1-\epsilon/3)}$$ 

\noindent where $\epsilon$ is the temperature- and wavelength-dependent
limb-darkening coefficient. In the Fourier domain, the convolution
product becomes an arithmethic product, and the Fourier transform of
$G(\lambda$) has the interesting property of having successive zeroes
at frequencies inversely proportional to $v \sin i$ (e.g., Dravins et
al. 1990), with the first zero occuring at
$\nu_1\simeq(2/3)(c/\lambda_o)\cdot(v\sin i)^{-1}$. Thus, even without
the knowledge of the intrinsic line profile, the projected stellar
velocity can be precisely derived from the location of the first and
subsequent zeroes in the Fourier transform of the observed
profile. This powerful technique has been most succesfully applied to
fast rotators ($v\sin i\geq$30~kms$^{-1}$) as their first zero occurs
in the well-sampled, high S/N low frequency Fourier domain. The
highest $v\sin i$ measured so far with this technique,
$\sim$600~kms$^{-1}$, was reported for an O-type star in the Large
Magellanic Cloud (Dufton et al. 2011). For a few bright stars, the
Fourier technique may even provide an estimate of surface latitudinal
differential rotation (Gray 1977; Reiners \& Schmitt 2002).  In
contrast, this method is not well suited to slowly rotating stars
($v\sin i\leq$20~kms$^{-1}$) whose first zero is usually lost in the
high frequency Fourier noise.

A more common method used to measure the rotation rate of slow
rotators is the cross-correlation analysis. Instead of measuring the
Doppler broadening of a single line profile, this method consists in 
cross-correlating the observed photospheric spectrum with either a
template spectrum of a star of similar effective temperature and
negliglible rotation (Tonry \& Davies 1979) (alternatively, a
non-rotating model spectrum can be used) or with a digital mask that
let light go through predefined wavelength ranges corresponding to the
location of major photospheric lines (Griffin 1967; Baranne et
al. 1979). The result of either process is a cross-correlation profile
or function (CCF) whose width is proportional to $v\sin i$ and whose
signal-to-noise ratio has been greatly enhanced thanks to the inclusion of
thousands of spectral lines in its computation. The relationship
between the CCF width and $v\sin i$ has to be properly calibrated
using stars with known rotation rates (Benz \& Mayor 1981, 1984;
Hartmann et al. 1986). Other applications of the cross-correlation
technique include the derivation of accurate radial velocities (CCF peak
location) and metallicity (CCF area). 

More sophisticated spectroscopic techniques have also been used to
measure rotation rates. The Doppler imaging technique (Vogt \& Penrod
1983) and the related Zeeman-Doppler imaging technique (Semel 1989;
Donati et al. 1997) both take advantage of the relationship existing
between the location of a feature at the surface of a rotating star
and its position within the line profile (Khokhlova 1976). As the star
rotates, the signatures of stellar spots (or magnetic components in
polarized light) move across the line profile and their monitoring
allows the reconstruction of surface brightness and/or magnetic
maps. The shape of the line profiles is thus periodically modulated by
surface inhomogeneities, and the modulation period provides a direct
measurement of the star's rotational period. Furthermore, the
latitudinal drift of spots on the stellar surface probes the
rotational period at different latitudes, thus yielding an estimate of
differential rotation at the stellar surface. Specifically, the
quantity $\Delta\Omega$ is derived by assuming a simplified solar-like
differential rotation law of the form:
\begin{equation}
\Omega(\theta)=\Omega_{eq} - \Delta\Omega \sin^2\theta
\end{equation}
where $\Omega_{eq}$ is the angular velocity at the stellar equator and
$\theta$ the latitude at the stellar surface. The relationships between
surface differential rotation on the one hand and effective
temperature, convective zone depth, and rotation rate on the
other, have been investigated for solar-type and lower mass stars by,
e.g., Barnes et al. (2005) and Marsden et al. (2011).
 
\subsection{Interferometry}

For relatively nearby stars, the stellar disk may be resolved by
interferometry (e.g. Kervella et al. 2004). In such a case, the
stellar oblateness, i.e., the decimal part of the ratio between the
equatorial to the polar radii can be measured. For rapidly-rotating
stars, the stellar oblateness can be quite substantial. According to
the Roche model for stellar equipotential surfaces, the
latitude-dependent radius of a fast-rotating star is given
by: $$R(\omega ,\theta) = {{3 R_{pole}}\over{\omega \sin \theta}}
\cdot cos[{{\pi + cos^{-1}(\omega \sin\theta)}\over{3}}]$$ where
$\theta$ is the co-latitude and $\omega=\Omega/\Omega_{crit}$ is the
ratio between the star's angular velocity and the critical velocity at which
centrifugal forces at the equator balance gravity,
$\Omega_{crit}=({2\over3})^{3\over 2}\cdot
({{GM}\over{R_p^3}})^{1\over 2} $, where $M$ is the stellar mass and
$R_p$ the polar radius, Ekstr\"om et al. 2008). For a star rotating at
critical velocity, the equatorial radius is 1.5 times larger than the
polar radius, yielding a stellar oblateness of 0.5. A stellar
oblateness with values up to 0.35 has been measured by interferometry
for a handful of massive stars rotating close to break-up velocity
(cf. van Belle 2012). 

Whenever interferometry can provide a fully reconstructed surface
brightness map for a rapidly rotating star, the gravity darkening
effect can be directly observed. von Zeipel (1924)'s theorem relates
radiative flux to surface gravity and thus predicts that a star
rotating close to break-up will be brighter at the pole than at the
equator. This has actually been observed in Altair ($v\sin
i=$240~kms$^{-1}$) by Monnier et al. (2007) who modeled the surface
brightness map of this star to derive both the inclination angle of
the rotational axis on the line of sight and its position angle on the
sky plane. This illustrates the complementary power of interferometry
compared to spectroscopy, as the former delivers the orientation of
the angular momentum vector in space while the latter yields its
(projected) modulus.

In fact, interferometry and spectroscopy can be combined, a technique
called spectro-interferometry, to measure the position angle of the
rotational axis of stars whose surface is not fully spatially
resolved. The method consists in measuring the position of the star's
photocenter across a spectral line. Each velocity channel within the
line profile corresponds to one isorotation stripe at the stellar
surface, parallel to the rotational axis. For instance, the far
redward wing of the line profile spatially coincides with the limb of
the receding hemisphere. As the photocenter is recorded across the
line in successive velocity channels, its location slightly moves on the sky
plane in a direction perpendicular to the rotational axis. Using
several interferometric baselines with different orientations, the
direction of the projected rotational axis can thus be derived.  This
challenging technique was successfully applied by Le Bouquin et
al. (2009) to demonstrate that the position angle of Fomalhaut's
rotational axis is perpendicular to the major axis of its
planet-hosting debris disk.

\subsection{Photometry}

The oldest method used to measure stellar rotation consists in monitoring
the visibility of magnetic spots on the stellar surface. In the
Western world, Galileo Galilei was amongst the first observers of the
early 17$^{th}$ century to provide an estimate of the Sun's rotational
period by observing the sunspots being carried across the stellar disk
by the star's rotation (Casas et al. 2006). When the stellar surface
is not resolved, starspots still modulate the star's luminosity in a
periodic way. Hence, the recording of the photometric light curve and
the detection of a periodically modulated signal provide a direct
estimate of the star's rotational period $P_{rot}$. This technique has
the advantage over spectroscopy of yielding a measurement of the
stellar rotation rate that is free of geometric effects and is
straighforwardly converted to angular velocity, $\Omega=2 \pi /
P_{rot}$. However, this is at the expense of requiring intense
photometric monitoring over several rotational periods and applying
dedicated signal processing techniques in order to recover the
periodic component of the light curve that truly corresponds to the
star's rotational period (e.g. Irwin et al. 2009).

This technique recently flourished with the Corot and Kepler
satellites that acquired continuous stellar light curves of exquisite
precision over timescales of months to years (e.g. Affer et al. 2012;
McQuillan et al. 2013). The large number of rotational cycles recorded
by these light curves allows not only the stellar rotational period to
be derived with extreme accuracy but also to detect latitudinal
differential rotation by traking spots located at different latitudes
that have slightly different rotational periods (e.g. Mosser et
al. 2009). The application of this technique is obviously most suited
to magnetically active stars that exhibit starspots at their surface,
i.e, usually solar-type and lower mass stars with a spectral type from
late-F to M, extending even to brown dwarfs (e.g. Herbst et al. 2007).

\subsection{Sismology}

The oscillation spectrum of a star encodes its rotational properties
from the surface down to the deep interior. To first order, the oscillation
frequencies of radial order $n$, degree $l$, and azimuthal order $m$ of
a rotating star are related to the same frequencies in a non-rotating
star by: $$ \nu_{n,l,m}=\nu_{n,l}-m\nu_s$$ where $m\in
[-l,+l]$ and $\nu_s \simeq \nu_{rot} = \Omega/2\pi$ (Goupil et
al. 2004). Stellar rotation lifts the $m$-degeneracy of the
oscillation modes of a non-rotating star by producing a rotational
splitting whose amplitude is directly proportional to angular velocity. In
case of uniform rotation, the rotational splitting of the modes 
provides a direct measurement of surface angular velocity. For more
complex rotational profiles, modeling the frequency splitting with
rotating stellar models offers a way to estimate the internal rotation
profile of the star. This technique was first applied to the Sun to
recover the latitudinal and radial variations of the solar rotation
rate through the convective envelope and down into the radiative core
(e.g., Schou et al. 1998). More recently, a similar approach based on
the analysis of the rotational splitting of mixed pressure and gravity
modes allowed Deheuvels et al. (2012) to probe the internal rotation
profile of a low-mass giant evolving off the main sequence, thus
revealing a rapidly rotating inner core.

Additional information can be retrieved from the amplitude of the
rotationally splitted modes. For instance, the ratio of the amplitudes
of the $m=0,\pm 1$ mode components depends on the inclination of the
rotational axis on the line of sight (Gizon \& Solanki 2003).  A first
application of this technique has recently allowed Chaplin et
al. (2013) to demonstrate that the rotational axis of 2 transiting
exoplanet hosts detected by Kepler is perpendicular to the orbital
plane of the planets.

\begin{table*}
\caption{Definitions of physical quantities related to stellar rotation.}
\label{def}
\begin{tabular}{|l|ll|l|}
\hline
\hline
{\bf Quantity} & {\bf Symbol} & {\bf Units} & {\bf Solar value}\\
\hline
Projected linear velocity & \vsini\ = V$_{eq}\sin i $ & km s$^{-1}$ &
$V_{eq,\odot}\simeq$ 1.9 km s$^{-1}$ \\
Rotational period & P$_{rot}$ = ${2\pi R_* \over V_{eq}}$ & days & P$_{eq,\odot}\simeq$ 26~d \\
Angular velocity & $\Omega_*= {V_{eq}\over R_*}={2\pi \over P_{rot}}$ & s$^{-1}$
& $\Omega_\odot$ = 2.8 10$^{-6}$ s$^{-1}$\\
& & & \\
Critical velocity & $V_{crit} = \sqrt{ {2\over 3} {G M_* \over
    R_*} }$
    & km s$^{-1}$ & $V_{crit,\odot}\simeq$ 360 km s$^{-1}$ \\
Latitudinal differential rotation & $\Omega(\theta)=\Omega_{eq} -
\Delta\Omega \sin^2\theta$ & s$^{-1}$ & $\Delta\Omega_\odot$ = 4.8
10$^{-7}$ s$^{-1}$\\
& & & \\
Moment of inertia & $I={8\pi\over 3}\int_0^{R_*} r^4\rho(r)dr$
& g cm$^2$ & $I_\odot$ = 6.4 10$^{53}$ g cm$^2$ \\
 Angular momentum & $J={8\pi\over 3}\int_0^{R_*} r^4\rho(r)\omega(r)dr$
& g cm$^2$ s$^{-1}$ & $J_\odot^\dagger$ = 1.8 10$^{48}$ g cm$^2$ s$^{-1}$ \\
 Specific angular momentum & $j=(J/M)$ & cm$^2$ s$^{-1}$ & $j_\odot$ =
 9 10$^{14}$ cm$^2$ s$^{-1}$\\
& & & \\
&\multicolumn{2}{c|}{N.B. If $\omega(r)=\Omega_*$,
  $J=I\Omega_*=k^2M_*R_*^2\Omega_*$,} & $k^2_{conv,\odot}$ = 0.008\\
& \multicolumn{2}{c|}{where $kR_*$ is the stellar radius of gyration$^\ddagger$}
& $k^2_{rad,\odot}$ = 0.061 \\
& & & \\
\hline
\end{tabular}
$^\dagger$ Pinto et al. (2011); $^\ddagger$ cf. Ruci\'nski (1988)
\end{table*}

\section{ The rotational properties of solar-type and lower mass
  stars}
\label{lowmass}

We review in this section the rotational properties of stars with a
mass less than 1.2M$_\odot$, from early studies to the most recent
determinations of rotational period distributions ranging from the
early pre-main sequence (PMS) to the end of the main sequence
(MS). The definition of physical quantities related to stellar
rotation that we use in this Section are summarized in
Table~\ref{def}. Solar values are listed for reference.

\subsection{Early studies and concepts}
\label{early}

\subsubsection{Rotation on the Main Sequence}

Kraft (1970) provided one of the first reviews on the rotational
properties of stars on the main sequence. The main characteristics of
the rotation rate distribution was a sharp break in velocity at a
spectral type around F4, i.e., around a mass of $\sim$1.2~M$_\odot$,
with more massive stars having mean rotation rates of order of
100-200~kms$^{-1}$, while lower mass stars had much lower rotational
velocities of order of a few \kms. The sharp decline of rotation rate
for stars with deep convective envelopes had readily been interpreted
by Schatzman (1962) as the result of angular momentum loss due to
magnetized winds.  In this framework, all stars were born with high
rotation rates, and only magnetically active stars with surface
convective envelopes would undergo strong braking as angular momentum
is removed from their surface by magnetized stellar winds. In
magnetically active stars, the ionised outflow remains coupled to the
magnetic field out to a distance where the magnetic tension becomes
unable to compensate for Coriolis force, i.e.: $$r\simeq
{{B^2}\over{16\pi\omega v\rho}}$$ where B is the magnetic field
intensity, $\omega$ the surface angular velocity, $v$ the poloidal
velocity of the wind flow, and $\rho$ its density. The magnetic lever
arm up to this radius yields angular momentum loss rates that are
orders of magnitude larger than in the absence of magnetically-coupled
winds. As shown by Weber \& Davis (1967), the angular momentum loss
rate can be expressed as: \begin{equation}{{dJ}\over{dt}} = {2\over 3} \Omega
\dot{M} r_A^2 = { J\over \tau_\textrm{w}} \label{jdot}\end{equation} where $\Omega$ is the stellar angular
velocity, $\dot{M}$ the mass-loss rate, $r_A$ the Alfv\'en radius, and
$\tau_\textrm{w}$ the braking timescale. For the Sun, the Alfv\'en radius is
about 30 times larger than the solar radius, which translates into a
braking timescale by the magnetized wind of order of 1~Gyr, i.e., short
enough to account for the slow rotation of the Sun on the mid-main
sequence.

A spectacular confirmation of this magnetic wind braking concept came
with one of the first studies of rotational evolution among main
sequence stars. Based on an earlier suggestion by Kraft (1967),
Skumanich (1972) used published measurements of the mean rotation rate
of solar-type stars in 2 young open clusters, the Pleiades and the
Hyades, and comparing them to the Sun's rotation, derived his famous
time-dependent velocity relationship $V_{eq}\propto t^{-1/2}$, for
ages between 0.1 and 5~Gyr. This relationship is indeed what is
asymptotically expected from the magnetic wind braking process, as
shown by Durney \& Latour (1978).  Combining the following expression
for mass-loss:
\begin{equation}
\dot{M}=-4\pi\rho_au_ar_a^2
\end{equation}
where $\rho_a$ and $u_a$ are the density and poloidal velocity of the
outflow at the Alfv\'en radius $r_a$, with the definition of Alfv\'en
velocity: 
\begin{equation}
B_a^2=4\pi\rho_au_a^2
\end{equation}
where $B_a$ is the stellar magnetic field at the Alfv\'en radius, and
with the condition of magnetic flux conservation: 
\begin{equation}
B_or_o^2=B_ar_a^2
\end{equation}
and replacing the expressions above in Eq.~\ref{jdot},
yields:\begin{equation}{{dJ}\over{dt}} = {{2\Omega}\over {3u_a}}
  (B_or_o^2)^2\label{jdotboro}\end{equation} Further assuming that the
poloidal velocity of the outflow at the Alfv\'en radius reaches the
escape velocity ($u_a=v_{esc}$) and that the stellar magnetic field is
powered by an internal dynamo process wich scales as
$B_o\propto\Omega$, finally yields:\begin{equation}
  {{dJ}\over{dt}}\propto\Omega^3 = I
  {{d\Omega}\over{dt}}\label{djdtomega}\end{equation} which
asymptotically integrates to $\Omega\propto t^{-1/2}$, i.e., the
Skumanich relationship. While extremely satisfying conceptually, this
derivation makes a number of symplifying assumptions including
spherically symmetric radial magnetic field and wind, thermally-driven
outflows, and linear dynamo relationship, none of which strictly apply
to active young stars (see Section 4).

As \vsini\ measurements accumulated in the mid-80's especially for
stars located close to or on the zero-age main sequence (ZAMS) at an
age of about 100~Myr, it became clear that, at these young ages, a
large dispersion of rotation rates exists for solar-type and lower
mass stars. Thus, Stauffer (1987) reported \vsini\ ranging from less
than 10~\kms\ up to more than 150~\kms\ for G and K-type stars in the
Alpha Persei (80~Myr) and Pleiades (120~Myr) young open clusters, at
the start of their main sequence evolution. Clearly, this unexpectedly
large scatter of rotation rates at ZAMS pointed to a rotational
evolution during the pre-main sequence that was far more complex than
envisioned from the Skumanich relationship on the main sequence.

\subsubsection{Rotation during the Pre-Main Sequence}
\label{pms}

Extrapolating the Skumanich relationship back in time to the pre-main
sequence (PMS), at an age of $\sim$1~Myr, would predict rotational
velocities of order of 200~\kms.  Additionally, if protostellar
collapse is dominated by gravity, one should expect protostars to
rotate close to their break-up velocity.  It therefore came as a
surprise when the first measurements of rotational velocities for
solar-mass PMS stars revealed that their rotation rate rarely exceeds
25~\kms, i.e., about a tenth of the break-up velocity (Vogel \& Kuhi
1981; Bouvier et al. 1986; Hartmann et al. 1986). Even deeply embedded
protostars appear to exhibit quite moderate rotation, with a mean
value of about 40~\kms\ (Covey et al. 2005). Clearly, significant
angular momentum loss must occur during protostellar collapse and/or
during the embedded protostellar phase of evolution to account for
such low rotation rates as the stars first appear in the HR diagram
(see Hennebelle, this volume; Belloche, this volume). Like on the main
sequence, higher mass PMS stars exhibit larger rotational velocities
than their lower-mass counterparts (Dahm et al. 2012). Most of the
so-called Herbig Ae-Be stars actually have similar velocities than
their MS counterpart, which suggests they lose little angular momentum
during the PMS, except for the precursors of the peculiar subgroup of
magnetic A and B stars (cf. Alecian et al. 2013).

The low rotation rates of PMS low-mass stars is even more surprising
when considering that they accrete high specific angular momentum
material from their circumstellar disk for a few million years
(Hernandez et al. 2007). As shown by Hartman \& Stauffer (1989), a
star accreting at a rate $\dot{M}$ from its disk will gain angular
momentum at a rate:
\begin{equation} 
{{dJ}\over{dt}} = \dot{M}  R_*^2 \Omega_{Kep}
\end{equation}
where $\Omega_{Kep}$ is the Keplerian velocity of the disk
material. It is then expected to spin up to an equatorial velocity of:
\begin{equation}
V_{eq} \simeq {{R_*^2\int{\dot{M} dt}} \over {I}} \cdot V_{br}
\end{equation}
where $I\simeq 0.2 M_*R_*^2$ is the stellar moment of inertia and
$V_{br}$ the break-up velocity. Thus, for a mass accretion rate of a
few 10$^{-8}$~\msunyr\ lasting for about 3~Myr, one expects the young
star to rotate at more than half the break-up velocity. Clearly, since
most low-mass PMS stars have much slower rotation rates, the accretion
of high angular momentum material from the disk must be balanced by a
process that efficiently removes angular momentum from the central
star.

Based on a physical process thought to be at work in compact
magnetized objects such as accreting neutron stars, K\"onigl (1991)
was first to suggest that the magnetic interaction between the inner
disk and a young magnetized star might provide a way to remove part of
the angular momentum gained from accretion. Shortly after, evidence
for a correlation between rotation rate and accretion was reported
(Bouvier et al. 1993; Edwards et al. 1993), with accreting young stars
rotating on average more {\it slowly} than non-accreting ones, thus
providing a strong support to K\"onigl's suggestion. Yet, more than 20
years later, the controversy is still very much alive as to whether
the magnetic star-disk interaction is efficient enough to counteract
the accretion-driven angular momentum gain in young stars (see
Ferreira, this volume). Also, even though a number of recent studies
appear to confirm the early evidence for a rotation-accretion
connection in young stars (e.g. Rebull et al. 2006; Cieza \& Baliber
2007; Cauley et al. 2012; Dahm et al. 2012; Affer et al. 2013), some
discrepant results have also been reported (e.g. Le~Blanc et
al. 2011). Thus, while there is a general consensus for young
accreting stars being somehow prevented from spinning up as they
evolve towards the main sequence (e.g., Rebull et al. 2004), the
underlying physical mechanism responsible for this behaviour is not
totally elucidated yet.

\subsection{Recent developments}

\begin{figure*}
\centering
\includegraphics[angle=-90,width=\hsize]{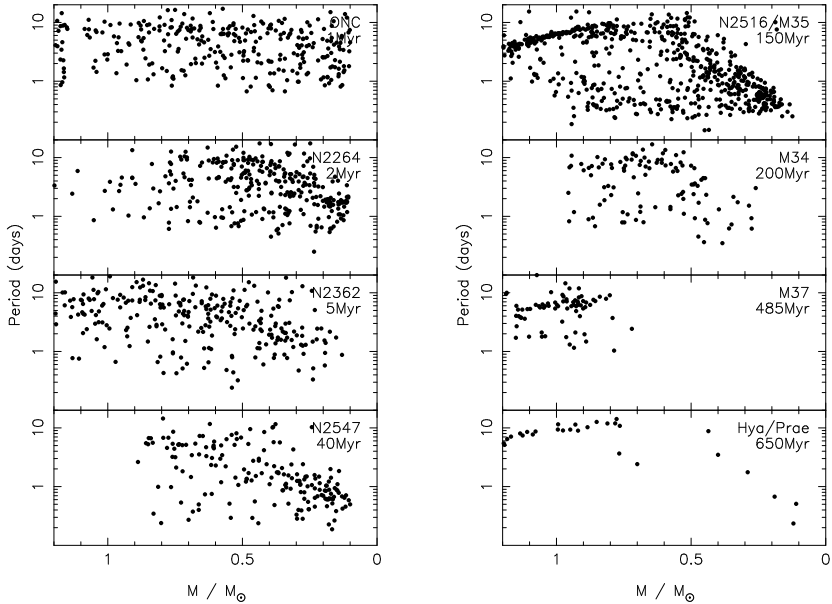}
\caption{Compilation of several thousands rotation periods for stars
    with masses M$\leq$1.2~M$_\odot$ in young clusters in the age
    range from 1~Myr to 0.6~Gyr. Plotted in each panel is rotation
    period as a function of stellar mass for a single cluster or from
    combined clusters when they have similar ages (e.g. NGC~2516/M35,
    Hyades/Praesepe). From Irwin \& Bouvier (2009). }
              \label{prot}%
\end{figure*}

While the early studies from the 60's to the 80's mostly focused on
the determination of projected rotational velocities, \vsini,
large-scale photometric monitoring campaigns started in the 90's that
provided complete rotational period distributions for thousands of
low-mass stars in the PMS and MS stages. Figure~\ref{prot} (from Irwin
\& Bouvier 2009, see references therein) illustrates a compilation of
some of these results. It shows how the distribution of rotational
periods evolves from the start of the PMS at about 1~Myr to the mid-MS
at 0.6~Gyr. A number of clear evolutionary trends emerge, which have
been confirmed by more recent studies. The initial distribution in the
Orion Nebulae Cluster at $\simeq$1~Myr is quite broad. It was found to
be bimodal for stars more massive than 0.3~M$_\odot$ with a slow
rotator peak with periods around 8~days and a fast rotator group with
periods around 2~days (Herbst et al. 2001). The peak of slow rotators
is usually attributed to PMS stars still interacting with their disk,
hence being prevented from spinning up, while the fast rotators are
thought to mainly consist of stars that have already dissipated their
circumstellar disks and therefore have started to spin up as they
evolve towards the ZAMS.  In constrast, the rotational period
distribution of very low-mass stars appears unimodal and skewed
towards faster rotators. As time progresses towards the ZAMS, which is
reached in about 40~Myr for a solar-mass star and 150~Myr for a
0.5~M$_\odot$ star, the period distributions evolve towards faster
rotation, especially in the low mass domain where rotational periods
at ZAMS converge to values less than 1~day. However, for stars more
massive than about 0.4~M$_\odot$, a large dispersion of rotation rates
remains up to the ZAMS. It is only later on the MS, by an age of about
0.5~Gyr, that all but the lowest mass stars are significantly braked,
to reach periods larger than about 10~days (Delorme et al. 2011;
Meibom et al. 2011), and exhibit a tight rotation-mass
relationship. At the low mass end, below 0.6~M$_\odot$, a significant
dispersion still subsists at that age (Scholz et al. 2011; Ag\"ueros
et al. 2011) and even beyond for most late-type field dwarfs
(McQuillan et al. 2013), lasting for perhaps as long as 10~Gyr for the
lowest mass stars (M$\leq$0.3~M$_\odot$; Irwin et al. 2011). Hence,
the spin down timescale on the main sequence significantly increases
towards lower mass objects, from a few 0.1~Gyr for solar-type and
low-mass stars up to a few Gyr for very low-mass stars (Delfosse et al
1998).

These recent studies have highlighted that the angular momentum (AM)
evolution of cool stars is strongly mass dependent, both during the
PMS and on the MS, as can be clearly seen from
Fig.~\ref{prot}. Schematically, solar-type and low-mass stars
(0.5-1.1~M$_\odot$) have a large initial dispersion of rotational
periods that subsists and even widens to the ZAMS, and is eventually
erased on the MS as all stars in this mass range are efficiently
braked on a timescale of a few 0.1~Gyr, thus yielding a well-defined
rotation-mass sequence with little scatter. The rotational convergence
of solar-type stars on the early MS has led to the developement of
gyrochronology, i.e., the measurement of stellar age from rotation
rate (e.g., Barnes 2003; Delorme et al. 2011; Epstein \& Pinsonneault
2012).  In contrast, very low-mass stars (M$\leq$0.3~M$_\odot$), while
also exhibiting some dispersion of rotation rates at the start of the
PMS evolution, seems to all converge towards fast rotation at ZAMS,
and resume building up a large rotational scatter on a timescale of a
few Gyr on the MS. This different behaviour is well illustrated by the
changing shape of the period-mass diagrams shown in Fig~\ref{prot} as
time goes by.

Going deeper into the mass spectrum, brown dwarfs' (BD's,
M$\leq$0.08~M$_\odot$) rotational properties seem to mimic and extend
those of very low-mass stars (Herbst et al. 2007;
Rodr\'{\i}guez-Ledesma et al. 2009), with no apparent rotational
discontinuity at the stellar/substellar boundary. As a group, they
tend to rotate faster than stars at all ages (Mohanty \& Basri
2003), with a median period of order of 15~hours at young ages, and
some indeed with rotational periods as short as a few hours, i.e.,
reaching close to the rotational break-up (Scholz \& Eisl\"offel 2004,
2005). Rapid rotation is still measured for evolved BD's at an age of
a few Gyr, which suggests that they suffer much weaker angular
momentum losses than stars (Reiners \& Basri 2008).

\section{Modeling the angular momentum evolution of cool stars}
\label{models}

The wealth of new data acquired since the mid-90's, now encompassing
several thousands of rotational periods measured for cool stars over
an age range covering from the start of the PMS to the late-MS
prompted renewed interest in the development of angular momentum
evolution models. While a review of all existing models and their
origin is far beyond the scope of this introductory course, we outline
in this section the main physical processes that are thought to drive the
rotational evolution of low-mass stars and how they are currently
implemented in parametrized models of angular momentum evolution.

\subsection{The physical processes behind rotational evolution} 

The rotational evolution of low-mass stars is believed to be dictated
by 3 main physical processes: star-disk interaction in the early PMS,
magnetized wind braking, and AM transport in the stellar interior. We
only briefly summarize these processes below, as they are reviewed in
much more detailed in other contributions to this volume.

\subsubsection{Star-disk interaction during the early PMS}
\label{sdi}

Camenzind (1990) and K\"onigl (1991) were first to suggest that the
low rotation rates of PMS stars may result from the magnetic star-disk
interaction. The picture envisioned at that time was inspired by the
Ghosh et al.'s (1977) model developed for accreting neutron stars
(e.g. Collier Cameron et al. 1995). While this model now does not seem to be
efficient enough to apply to young stars, a number of alternatives
have been proposed still relying on the inner disk interacting with a
strong stellar magnetosphere. Indeed, young stars are known to host
strong magnetic fields (cf. Donati, this volume) that are able to
disrupt the inner disk regions and channel the accretion flow onto the
star through magnetic funnels (cf. Bouvier et al. 2007 for a
review). Bessolaz et al. (2008) derived the following expression for
the magnetospheric truncation radius:
\begin{equation}
{{r_{tr}}\over{R_*}}\simeq 2 m_s^{2/7} \left({{B_*}\over{140G}}\right)^{4/7}
\left({{\dot{M}_a}\over{10^{-8}M_\odot yr^{-1}}}\right)^{-2/7}
\left({{M_*}\over{0.8 M_\odot}}\right)^{-1/7} \left({{R_*}\over{2 R_\odot}}\right)^{5/7}
\end{equation}
where $m_s$ is the sonic Mach number at the disc midplane, $B_*$ the
stellar magnetic field, and $\dot{M}_a$ the mass accretion rate. For
values of the parameters relevant to a young accreting solar-mass
system, the magnetospheric truncation radius is located a
few stellar radii above the stellar surface, a prediction borne out by
observations (e.g. Najita et al. 2007). This distance is of the same
order as the disk corotation radius, i.e., the radius at which the
keplerian angular velocity in the disk equals the star's angular
velocity: 
\begin{equation} 
  r_{co} = \left({{G M_*}\over{\Omega_*^2}}\right)^{1/3}
\end{equation} 
The net flux of angular momentum exchanged between the star and the
disk strongly depends upon whether the magnetospheric truncation
radius is located within or beyond the disk corotation
radius. Therefore, the changing magnetic topology of solar-type stars
evolving on their convective and radiative PMS tracks will most likely
impact their early rotational evolution (e.g. Gregory et al. 2012).

Within this general framework, various scenarios have been developed
to attempt to produce a negative net angular momentum torque onto the
star, so as to explain why young stars are slow rotators in spite of
both accretion and contraction. These include accretion-driven winds
(Matt \& Pudritz 2008), X-winds and their variants (Mohanty and Shu
2008; Ferreira et al. 2000) , and magnetospheric ejections (Zanni \&
Ferreira 2013). These models are discussed at length in J.~Ferreira's
contribution to this volume. Whether any of these processes is
actually able to counteract the spin up due to accretion and
contraction during the early PMS is, however, unsettled. Pending a
satisfactory model for PMS spin down, most current angular momentum
evolution models assume that accreting PMS stars evolve at constant
angular velocity (cf. Sect.~\ref{AMmodels}). 

\subsubsection{Rotational braking by magnetized winds}
\label{wind}

Starting from the general expression of angular momentum loss due to
magnetized stellar winds (Eq. 3.1 above), Kawaler (1988) worked out a
parametrized formulation that can be straightforwardly implemented in
evolutionary models. Following Mestel (1984), the AM loss rate is given by:
\begin{equation}
{\dot{J}_\textrm{w}} = {2\over 3} \dot{M} \Omega_*
 R_*^2 \left[\left({{r_A}\over{R_*}}\right)_{radial}\right]^n  
\end{equation}
where $\dot{M}$ is the mass loss rate, $\Omega_*$ the stellar angular
velocity, $R_*$ the stellar radius, $r_A$ the Alfv\'en radius, and the
exponent $n$ reflects the magnetic field geometry with $n=2$ for a
radial field and $n=3/7$ for a dipolar field. From Eq. 3.2, 3.3,
and 3.4 above, the expression of the Alfv\'en radius is given by:
\begin{equation}
\left({{r_A}\over{R_*}}\right)_{radial}^2 = {{B_*^2 R_*^2}\over{\dot{M} u_a}}
\end{equation}
where $B_*$ is the stellar magnetic field intensity, and $u_a$ the
flow velocity at the Alfv\'en radius.  Further assuming that the flow
velocity at the Alfv\'en radius is of order of the escape velocity,
and adding a dynamo relationship for the generation of the stellar
magnetic field of the form: 
\begin{equation}
B_* R_*^2 \propto \Omega_*^a
\end{equation}
where $a$ is the dynamo exponent, finally leads to:  
\begin{equation}
{\dot{J}_\textrm{w}} = -K_\textrm{w} \dot{M}^{1-(2n/3)} \Omega_*^{1+(4an/3)}
 R_*^{2-n} M_*^{-n/3}
\end{equation}
For an asumed linear dynamo relationship ($a$=1), and a magnetic
topology intermediate (in some sense) between a radial and a dipolar
field with $n$=1.5, Eq.4.6 simplifies to:
\begin{equation}
{\dot{J}_\textrm{w}} = -K_\textrm{w} \Omega_*^3 R_*^{1/2} M_*^{-1/2}
\end{equation}
which is easily implement in an evolutionary model. Direct application
of this prescription, however, proved to produce too strong braking
for fast rotators compared to observations (Stauffer \& Hartmann
1987). Most models have therefore adopted a variant of Kawaler's
prescription, first proposed by Chaboyer et al. (1995), that assumes
that the dynamo saturates ($a=0$) above some angular velocity
$\omega_{sat}$, i.e.:
\begin{equation}
{\dot{J}_\textrm{w}} = 
\left\{
\begin{array}{l l}
-K_\textrm{w} \Omega_*^3 R_*^{1/2} M_*^{-1/2} &  \Omega\leq \omega_{sat} \\
-K_\textrm{w} \Omega_* \omega_{sat}^2 R_*^{1/2} M_*^{-1/2} &
\Omega_*>\omega_{sat}
\end{array}
\right.
\end{equation}

The shallower slope of the rotation-dependent angular momentum loss at
high rotation, i.e., $\dot{J}_\textrm{w}\propto\Omega$ instead of
$\dot{J}_\textrm{w}\propto\Omega^3$, provides a better agreement with
the observation of very fast rotators at the ZAMS. Magnetic field
measurements suggest that dynamo saturation occurs at a fixed Rossby
number $R_0\simeq 0.1$ in cool stars (Reiners et al. 2009; Wright et
al. 2011), with $R_0=2\pi (\omega \tau_c)^{-1}$ where $\tau_c$ is the
turnover convective time. As $\tau_c$ lengthens towards lower mass
stars, $w_{sat}$ is expected to decrease with mass, i.e., lower mass
stars suffer less angular momentum loss than solar-type ones. This
Rossby scaling thus naturally accounts for the longer spin down
timescale of lower mass stars on the main sequence (Krishnamurthi et
al. 1997; Bouvier et al. 1997). However, Sills et al. (2000) showed
that very low-mass stars (M$\leq$0.4~M$_\odot$) experience much less spin
down than the extrapolation of the Rossby scaling to very low masses
would predict. Recently, Reiners \& Mohanty (2012) proposed a
modification to Kawaler's prescription, based on a more
physically-funded dynamo relationship, that appears to alleviate this
issue.

Recent MHD numerical simulations of stellar winds have considerably
improved our understanding of wind-driven angular momentum loss
(e.g. Aarnio et al. 2012; Vidotto et al. 2009, 2011).  Based on
2D numerical simulations of MHD winds originating from stars with a dipolar
magnetic field, Matt et al. (2012) derived the following expression for the
Alfv\'en radius:
\begin{equation}
\label{matt}
{r_A \over R_*} = K_1 \left[   {\Upsilon \over {(K_2^2 + 0.5f^2)^{1/2}}}  \right]^{m}
\end{equation}
where $f$ is the ratio of the stellar rotation rate to the break-up
velocity and  
\begin{equation}
\Upsilon = {{B_*^2 R_*^2}\over {\dot{M}_\textrm{w} v_{esc}}}
\end{equation} 
where $B_*$ is the magnetic field strength at the stellar equator,
$\dot{M}_\textrm{w}$ the wind mass loss rate, and $v_{esc}$ the escape
velocity. The value of the constants appearing in Eq.~\ref{matt} are derived
from numerical simulations that explore the parameter space, yielding
$K_1=1.3$, $K_2=0.0506$ and $m=0.2177$. Provided the stellar magnetic
field can be tied to the angular velocity through a dynamo
prescription, and the wind mass loss rate can be computed as a
function of stellar rotation and other fundamental stellar parameters
(e.g. Cranmer \& Saar 2011), the expression given by Eq.~\ref{matt} for
the Alfv\'en radius can be implemented in Eq.~\ref{jdot} to compute the
amount of wind-driven angular momentum losses during stellar
evolution. Models using this new prescription for AM losses are
illustrated below (cf. Sect.~\ref{AMmodels}).

\subsubsection{Angular momentum transport in stellar interiors}
\label{decoupling}

As angular momentum is carried away by stellar winds at the stellar
surface, several mechanisms may operate to redistribute angular
momentum in the stellar interior. These range from various classes of
hydrodynamical instabilites (e.g. Lagarde et al. 2012, see also the
contributions of Palacios and Rieutord in this volume), magnetic
fields (e.g. Eggenberger et al. 2005), and gravity waves (Talon \&
Charbonnel 2008; Charbonnel et al. 2013; see also Mathis, this
volume). The recent report of rapidly rotating cores in red giants
from asterosismology (e.g. Mosser et al. 2012) and the discrepancy
between the measured angular velocity gradient and model expectations
indicate that angular momentum transport mechanisms in stellar
interiors are still not totally elucidated (e.g. Eggenberger et
al. 2012).

Lacking a detailed physical modeling of the processes involved,
MacGregor \& Brenner (1991) introduced a parametrized prescription for
angular momentum transport between the radiative core and the
convective envelope. Each region is considered as rotating uniformely
but not necessarily at the same rate, as the convective envelope is
slowed down. They assumed that AM transport processes would act to
erase angular velocity gradients at the boundary between the radiative
core and the convective envelope (the tachocline, cf. Spiegel \& Zahn
1992), on a timescale $\tau_{ce}$, the so-called core-envelope
coupling timescale. To reach a state of uniform rotation on a
timescale $\tau_{ce}$ throughout the star, a quantity $\Delta J$ of
angular momentum has to be exchanged between the radiative core and
the convective envelope, with:
\begin{equation}
\Delta J = {{I_{conv} J_{core} - I_{core} J_{conv}}\over{I_{core}+I_{conv}}}
\end{equation}
 where $I_{conv}$, $I_{core}$ and $J_{conv}$, $J_{core}$ are the
 moment of inertia and the angular momentum content of the convective
 envelope and the radiative core, respectively. Then, the angular
 momentum evolution of the radiative core and the convective envelope
 can be written as:
\begin{equation}
{dJ_{core}\over dt} = - {\Delta J \over \tau_{ce}} 
\end{equation} 
and
\begin{equation}
{dJ_{conv}\over dt} = {\Delta J \over \tau_{ce}} - {J_{conv} \over \tau_J}
\end{equation}
where $\tau_J$ is the wind braking timescale\footnote{Note that during
  the pre-main sequence, additional terms enter the above equations
  governing the evolution of the core and envelope angular momenta as
  the radiative core develops in the initially fully convective star
  (cf. Allain 1998).}.  This prescription has been used in two-zone
angular momentum evolution models (e.g. Allain 1998) that provide some
insight into the value of $\tau_{ce}$ and its dependence upon rotation
rate and stellar parameters. A short coupling timescale corresponds to
an efficient AM redistribution and leads to solid-body rotation, while
a long $\tau_{ce}$ allows for strong angular velocity gradients to
develop at the tachocline as the star evolves. Thus, this
parametrization offers some empirical guidance to identify the actual
underlying physical mechanisms at work for angular momentum transport
in stellar interiors based on the timescales involved.

\subsection{Parametrized models of angular momentum evolution}
\label{AMmodels}

Angular momentum evolution models have been developed in an attempt
to reproduce the run of surface rotation as a function of time, as
derived from observations for solar-type stars, low-mass and very
low-mass stars. In this section, we illustrate a class of
semi-empirical models that use parametrized prescriptions to implement
the physical processes described in the previous section.

\subsubsection{Solar-type stars}
\begin{figure*}
\centering
\includegraphics[angle=-90,width=\hsize]{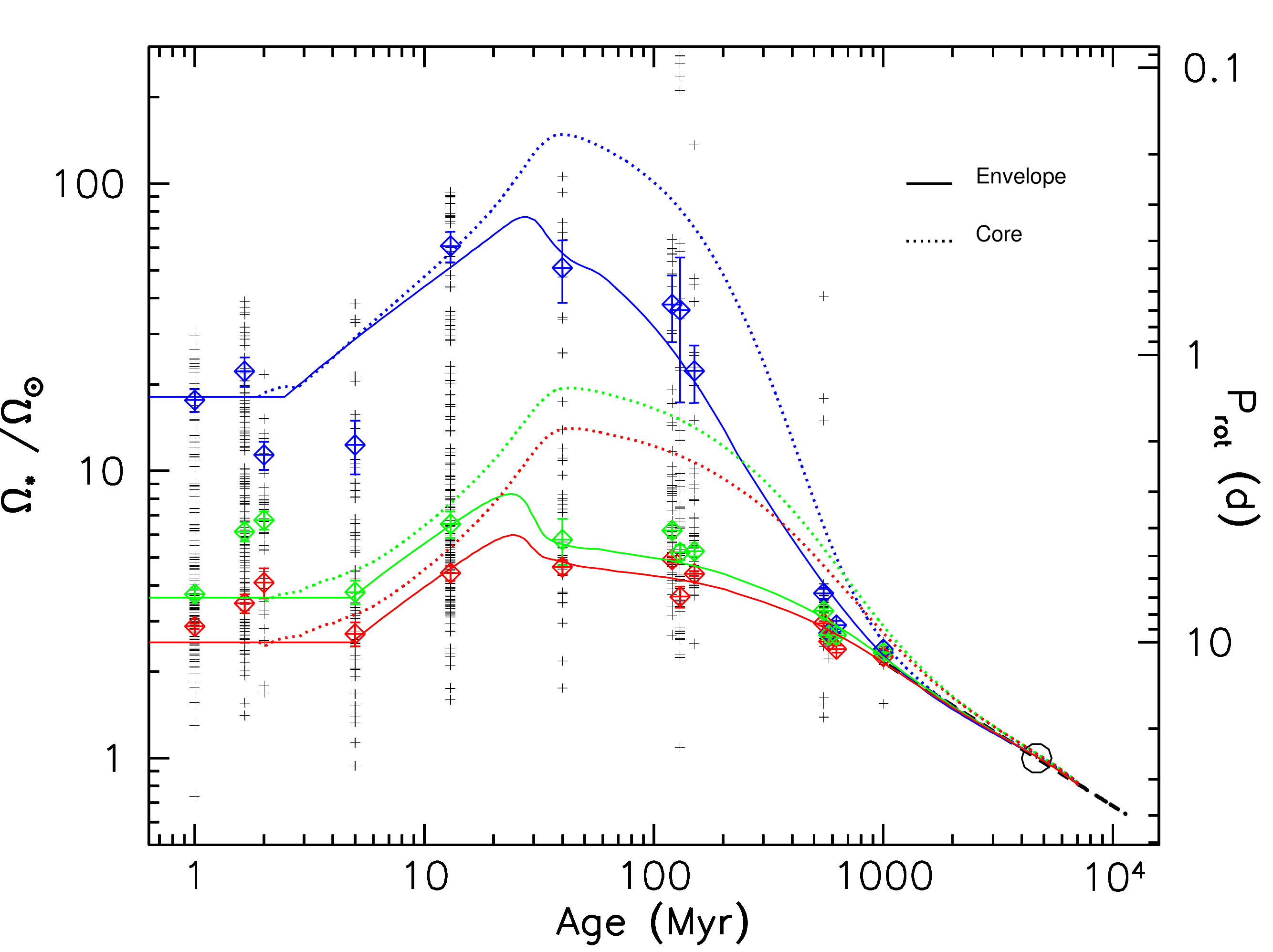}
\caption{The rotational angular velocity of solar-type stars is
  plotted as a function of age. The left y-axis is labelled with
  angular velocity scaled to the angular velocity of the present Sun
  while the right y-axis is labelled with rotational period in
  days. On the x-axis the age is given in Myr. {\it Observations:} The
  black crosses shown at various age steps are the rotational periods
  measured for solar-type stars in star forming regions and young open
  clusters over the age range 1~Myr-1~Gyr.  The associated red, green,
  and blue diamonds represent the 25, 50, and 90$^{th}$ percentiles of
  the observed rotational distributions. The open circle at 4.56~Gyr
  is the angular velocity of the present Sun.  {\it Models:} The
  angular velocity of the convective envelope (solid line) and of the
  radiative core (dashed lines) is shown as a function of time for
  slow (red), median (green), and fast (blue) rotator models, with
  initial periods of 10.0, 7.0, and 1.4~days, respectively. The dashed
  black line at the age of the Sun illustrates the asymptotic
  Skumanich's relationship, $\Omega\propto t^{-1/2}$. From Gallet \&
  Bouvier (2013).}
\label{gallet}%
\end{figure*}

Figure~\ref{gallet} (from Gallet \& Bouvier 2013) illustrates the
observed and modeled angular momentum evolution of solar-type stars,
in the mass range 0.9-1.1~M$_\odot$, from the start of the PMS at
1~Myr to the age of the Sun. The rotational distributions of
solar-type stars are shown at various time steps corresponding to the
age of the star forming regions and young open clusters to which they
belong (see Fig.2). Three models are shown, which start with initial periods of
10, 7, and 1.4~days, corresponding to slow, median, and fast
rotators. The models assume constant angular velocity during the
star-disk interaction phase in the early PMS (cf.~\ref{sdi}),
implement the Matt et al. (2012) wind braking prescription
(cf.~\ref{wind}), as well as core-envelope decoupling
(cf.~\ref{decoupling}). The free parameters of the models are the
initial periods, scaled to fit the rotational distributions of the
earliest clusters, the star-disk interaction timescale $\tau_d$ during
which the angular velocity is held constant at its initial value, the
core-envelope coupling timescale $\tau_{ce}$, and the calibration
constant $K_1$ for wind-driven AM losses. The latter is fixed by the
requirement to fit the Sun's angular velocity at the Sun's age. These
parameters are varied until a reasonable agreement with observations
is obtained. In this case, the slow, median, and fast rotator models
aim at reproducing the 25, 50, and 90$^{th}$ percentiles of the
observed rotational distributions and their evolution from the early
PMS to the age of the Sun.

The models provide a number of insights into the physical processes at
work. The star-disk interaction lasts for a few Myr in the early PMS,
and possibly longer for slow rotators ($\tau_d\simeq$5~Myr) than for
fast ones ($\tau_d\simeq$2.5~Myr). As the disk dissipates, the star
begins to spin up as it contracts towards the ZAMS. The models then
suggest much longer core-envelope coupling timescales for slow
rotators ($\tau_{ce}\simeq$30~Myr) than for fast ones
($\tau_{ce}\simeq$12~Myr). Hence, once they have reached the ZAMS,
slow rotators exhibit much lower surface velocities than fast rotators
but significantly larger angular velocity gradients at the
tachocline. Indeed, most of the initial angular momentum is hidden in
the core of the slow rotators at ZAMS. As they evolve on the early MS,
wind braking eventually leads to the convergence of rotation rates for
all models by an age of $\simeq$1~Gyr, to asymptotically reach the
Skumanich's relationship. These models thus clearly illustrate the
different rotational histories solar-type stars may experience,
depending mostly on their initial period and disk lifetime. In turn,
the specific rotational history a star undergoes may strongly impact
on its properties, such as lithium content, even long after rotational
convergence has taken place (cf. Bouvier 2008; Randich 2010).

\begin{figure*}
\centering
\includegraphics[angle=0,width=\hsize]{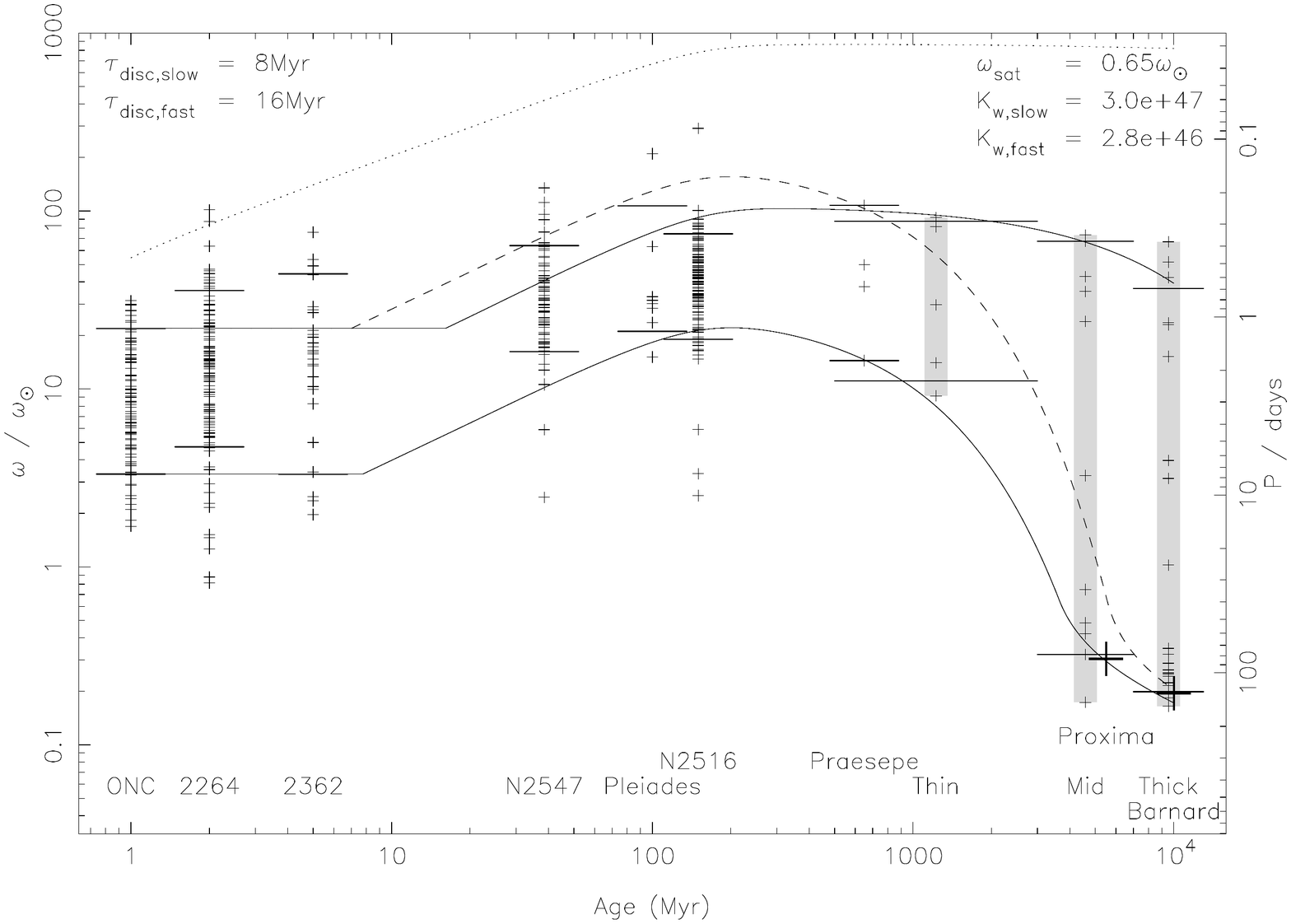}
\caption{The rotational angular velocity of very low-mass stars
  (0.1-0.35~M$_\odot$) is plotted as a function of age. The left
  y-axis is labelled with angular velocity scaled to the angular
  velocity of the present Sun while the right y-axis is labelled with
  rotational period in days. On the x-axis the age is given in
  Myr. {\it Observations:} The black crosses shown at various age
  steps are the rotational periods measured for very low-mass stars in
  star forming regions, young open clusters, and in the field over the
  age range 1~Myr-10~Gyr.  Short horizontal lines show the 10$^{th}$
  and 90$^{th}$ percentiles of the angular velocity distributions at a
  given age, used to characterize the slow and fast rotators,
  respectively. {\it Models:} The solid curves show rotational
  evolution models for 0.25~M$_\odot$ stars, fit to the percentiles,
  with the upper curve for the rapid rotators (with parameters
  $\tau_{d,fast}$ and K$_{\textrm{w},fast}$) and the lower curve for
  the slow rotators (with parameters $\tau_{d,slow}$ and
  K$_{\textrm{w},slow}$). Note the factor of 10 difference between
  K$_{\textrm{w},fast}$ and K$_{\textrm{w},slow}$. The dashed curve
  shows the result for the rapid rotators if the wind parameter
  K$_{\textrm{w},fast}$ is assumed to be the same as for the slow
  rotators rather than allowing it to vary. The dotted curve shows the
  break-up limit.  From Irwin et al. (2011).}
 \label{irwin}%
\end{figure*}

The models discussed above describe the rotational evolution of single
stars while many cool stars belong to multiple stellar systems. In
short period binaries ($P_{orb}\leq$12~days), tidal interaction will
enforce synchronization betweeen the orbital and rotational period
(Zahn 1977). Clearly, the rotational evolution of the components of
such systems is totally different from that of single stars, and rapid
rotation is usually maintained over the whole main sequence (Zahn \&
Bouchet 1989) and even beyond, like in, e.g., the magnetically-active
rapidly-rotating RS CVn systems. However, the fraction of such tight,
synchronized systems among solar-type stars is low, of order of 3\%
(Raghavan et al. 2010), so that tidal effects are unlikely to play a
major role in the angular momentum evolution of most cool stars.

Presumably much more frequent is the occurrence of planetary systems
around solar-type and low-mass stars (e.g.  Mayor et al. 2011; Bonfils
et al. 2013). The frequency of hot Jupiters, i.e., massive planets
close enough to their host star to have a significant tidal
influence\footnote{The magnetospheric interaction between the star and
  a hot Jupiter may also impact the stellar spin rate 
  (cf. e.g. Lanza 2010; Cohen et al. 2010)} (cf. Dobbs-Dixon et
al. 2004), is a mere 1\% around FGK stars (e.g. Wright et
al. 2012). However, there is mounting evidence that the planetary
formation process is quite dynamic, with gravitational interactions
taking place between forming and/or migrating planets (Albrecht et
al. 2012). This may lead to planet scattering and even planet
engulfment by the host star. The impact of such catastrophic events
onto the angular momentum evolution of planet-bearing stars has been
investigated by Bolmont et al. (2012) who showed it could be quite
significant both during the PMS and on the main sequence.

\subsubsection{Very low-mass stars}

Models similar to those described above for solar-type stars have been
shown to apply to lower mass stars, at least down to the fully
convective boundary ($\simeq$0.3~M$_\odot$), with the core-envelope
coupling timescale apparently lengthening as the convective envelope
thickens (e.g., Irwin et al. 2008). In the fully convective regime,
i.e., below 0.3~M$_\odot$, models ought to be simpler as the
core-envelope decoupling assumption becomes irrelevant and uniform
rotation is usually assumed instead throughout the star. Yet, the
rotational evolution of very low-mass stars actually appears more
complex than that of their more massive counterparts and still
challenges current models. Rotational period measurements for field
M-dwarfs show a bimodal distribution with a peak of fast rotators in
the period range 0.2-10~days, and a peak of slow rotators with
rotational periods ranging from 30~days to at least 150~days (Irwin et
al. 2011). Most of the slow rotators appear to be thick disk members,
i.e., they are on average older than the fast ones that are
kinematically associated to the thin disk. The apparent bimodality
coud thus simply result from a longer spin down timescale of order of
a few Gyr, as advocated by Reiners \& Mohanty (2012).

However, as shown in Figure~\ref{irwin} (from Irwin et al. 2011), this
bimodality may not be easily explained for field stars at an age of
several Gyr. It is seen that the large dispersion of rotation rates
observed at late ages for very low-mass stars requires drastically
different model assumptions. Specifically, for a given model mass
(0.25~M$_\odot$ in Fig.~\ref{irwin}), the calibration of the
wind-driven angular momentum loss rate has to differ by one order of
magnitude between slow and fast rotators (Irwin et al. 2011). Why does
a fraction of very low-mass stars remain fast rotators over nearly
10~Gyr while another fraction is slowed down on a timescale of only a
few Gyr is currently unclear. A promising direction to better
understand the rotational evolution of very low mass stars is the
recently reported evidence for a bimodality in their magnetic
properties. Based on spectropolarimetric measurements of the magnetic
topology of late M dwarfs (Morin et al. 2010), Gastine et al. (2013)
have suggested that a bistable dynamo operates in fully convective
stars, which results in two contrasting magnetic topologies: either
strong axisymmetric dipolar fields or weak multipolar fields. Whether
the different magnetic topologies encountered among M dwarfs is at the
origin of their rotational dispersion at late ages remains to be
assessed.

\section{The rotational properties of massive and intermediate-mass
  stars}
\label{highmass}

As outlined in Sect.~\ref{early}, stars more massive than
1.2~M$_\odot$ have significantly larger rotation rates than solar-type
and low-mass stars. In comparison to low-mass stars, more massive
stars have higher initial angular momenta, shorter contraction
timescales to the ZAMS and shorter evolution timescales on the MS,
they lack deep convective envelopes and strong magnetic fields (except
for peculiar sub-classes, such as Ap-Bp stars), and drive dense
radiative winds. For all these reasons, their rotational evolution is
expected to be quite different from that of their low-mass
counterparts. We briefly review in the next sections the rotational
properties of massive and intermediate-mass stars.

\subsection{Massive stars (4-15~M$_\odot$)}

Braganca et al. (2012) have recently summarized the rotational
distributions of 350 nearby O9-B6 stars in the Galactic disk from
\vsini measurements. After correcting for projection effect (i.e.,
$<V_{eq}> = {4\over \pi} <V\sin i>$, cf. Gaig\'e 1993), they find that
the mean equatorial velocity is of order of 125~\kms and relatively
uniform over the whole mass range they probe. A similar study was
performed by Huang \& Gies (2006) for 496 O9-B9 stars belonging to 19
young open clusters yielding an average equatorial velocity of
190~\kms, i.e., significantly higher than the mean rotation rate of
massive field stars. Specifically, the difference in mean velocity
between cluster and field stars at high masses stems from the much
smaller fraction of slow rotators observed in clusters, while the two
populations have similar \vsini\ distributions above $\simeq$100~\kms.
Even though cluster members are on average younger than field dwarfs,
Meynet \& Maeder (2000) rotating evolutionary models predict only
modest spin down for massive stars on the main sequence, with a
braking rate of order of 15-20\% for 9-12~M$_\odot$ stars.  Indeed,
the comparison of massive field dwarfs and cluster members over the
same age range (12-15~Myr) still result in differing average
velocities, thus suggestive of an intrinsic rather than an
evolutionary effect (Strom et al. 2005). Wolff et al. (2007) further
confirmed that massive stars formed in high-density regions, e.g. rich
clusters, lack the numerous slow rotators seen for stars of similar
masses in low-density regions and the field. These authors suggested
that the density-dependent rotational distribution observed for
massive stars may reflect a combination of initial conditions, e.g.,
higher turbulence in massive proto-clusters yielding larger initial
angular momenta, and environmental conditions, where the stronger
ambient UV flux from O-type stars in rich clusters may shorten the
disk lifetimes, thus minimizing the braking efficiency of the 
star-disk interaction during the early angular momentum evolution of
massive stars. Yet, it is unclear whether the disk locking scenario
discussed above for low-mass pre-main sequence stars does apply to
more massive stars (cf. Rosen et al. 2012). Nevertheless, regardless
of the actual physical processes at work, the observed relationship
between the shape of the rotational distributions of massive stars and
the specific properties of their birthplace seems to indicate that
initial conditions have a long-lasting impact on their rotational
properties.

\subsection{Intermediate-mass stars (1.3-4~M$_\odot$)}
    
As mentionned in previous sections (see \ref{pms}), intermediate-mass
PMS stars, the so-called Herbig Ae-Be stars, have on average much
higher rotational velocities than their lower mass T Tauri
counterparts at an age of a few Myr. Wolff et al. (2004) investigated
the rotational evolution of intermediate-mass PMS stars
(1.3-2~M$_\odot$) as they evolve from convective to radiative tracks
towards the ZAMS. Comparing the measured velocities on convective
tracks to those of stars landing on the ZAMS over the same mass range,
they concluded that angular momentum is conserved in spherical shells
within the star, i.e., that the PMS spin up is directly proportional
to the contraction of the stellar radius. The lack of angular momentum
redistribution in the predominantly radiative interiors of
intermediate-mass PMS stars would then imply that they develop a large
degree of radial differential rotation from the center to the surface
as they approach the ZAMS.

Further insight into the rotational properties and evolution of
intermediate-mass stars is provided by the large-scale study of Zorec
\& Royer (2012) who reported \vsini\ measurements for 2,014 B6- to
F2-type stars. By tracing the evolution of rotation as a function of
age, they confirm that intermediate-mass stars seem to evolve on the
main sequence as differential rotators. Striking differences between
the rotational distribution of 1.6-2.4~M$_\odot$ stars and that of
2.4-3.8~M$_\odot$ stars, the former being unimodal while the latter is
bimodal, remain to be understood.  Similarly, the complex rotational
behaviour of stars over this mass range as they evolve onto the MS,
with an apparent spin up during the first half of the MS evolution
followed by a significant spin down during the second half, represents
a real challenge for angular momentum evolution models.

A specific sub-group of intermediate-mas stars, the magnetic Ap-Bp
stars host surface magnetic fields of a few kG to a few 10~kG. This
sub-group represents about 5-10\% of the population and is known to
exhibit systematic lower velocities that their non-magnetic
counteparts (Abt \& Morrell 1995).  Alecian et al. (2012) showed that
their precursors, i.e., the magnetic Herbig Ae-Be stars, already are
slower rotators that non-magnetic intermediate-mass PMS stars,
indicating that magnetic braking is already efficient during the PMS
for this particuliar subgroup. As the spin down continues on the MS,
Ap stars can reach very slow rotation indeed, with the longest
rotational period ever reported amounting to 77$\pm$10 years (Leroy et
al. 1994).

\section{Conclusion}

The last decade has seen tremendous progress in the characterization
of the rotational properties of stars at various stages of evolution
and over the whole mass range from brown dwarfs to the most massive
objects. These new observational results bring formidable constraints to
the development of angular momentum evolution models. While the
dominant processes thought to dictate the rotational evolution of
stars are probably identified, much remains to be done to understand
their detailed physics and their respective roles. The confrontation
between models and observations, though much improved in recent years,
still indicate a number of shortcomings related to transport processes
in radiative interiors, the physics of stellar winds, and the
interaction between the star and its environment. Major advances are
expected to arise from multi-dimensional numerical simulations of
stellar interiors and stellar atmospheres, which will hopefully
provide new clues to the elusive physical processes that govern the
rotational evolution of stars from their birth to the last stages of
their evolution.

\begin{acknowledgements}
  It is a pleasure to thank the organisers of the Evry Schatzman
  School for a very enjoyable week, the friendly atmosphere, and the
  unexpected celebration event during the school. I would also like to
  thank Florian Gallet and Jonathan Irwin for providing Fig.3 and 4,
  respectively, of this contribution, Jean-Baptiste Le Bouquin for
  interesting discussions on interferometry, Rafael Garcia and
  J\'er\^ome Ballot for guidance in asterosismology, Juan Zorec for
  providing useful references on the rotation of intermediate-mass
  stars, Gaspard Duch\^ene for help on binary statistics, Gauthier
  Mathys for discussions on Ap stars, and Nad\`ege Meunier for an
  historical review of the first measurements of the Sun's rotation,
  whose paternity remains controversial.
\end{acknowledgements}


\end{document}